\documentstyle[aps,twocolumn,floats]{revtex}

\def \Rb {\hbox {Rb}}
\def \Cs {\hbox {Cs}}
\def \Ak {\hbox {Ak}}
\def \Tr {\hbox {Tr}}
\def \bzeta \mbox{\boldmath{ $\zeta$ }}

\begin{document}
%\draft
\title{Spin Relaxation Resonances Due to the Spin-Axis
        Interaction in Dense Rubidium and Cesium Vapor}
\author{C. J. Erickson, D. Levron\thanks{Permanent Address: Nuclear Research
Center, Beer Sheva, Israel.}, W. Happer}
\address{Joseph Henry Laboratory, Physics Department, Princeton University,
Princeton, New Jersey, 08544}
\author{S. Kadlecek, B. Chann, L. W. Anderson, T. G. Walker}
\address{Department of Physics, University of Wisconsin-Madison, Madison,
Wisconsin, 53706}
\date{\today}
\maketitle

\begin{abstract}
Resonances in the magnetic decoupling curves for
the spin relaxation of dense alkali-metal vapors prove that
much of the relaxation is due to the spin-axis interaction
in triplet dimers. Initial estimates  of the spin-axis coupling
coefficients  for the dimers (likely accurate to a factor of 2) are
$|\lambda|/h = 290$ MHz for Rb;
$2500$ MHz for Cs.
\end{abstract}

\bigskip

Hot, dense alkali-metal vapors are used to polarize the nuclear
spins of noble gases, especially $^3$He and $^{129}$Xe, by spin
exchange optical pumping\cite{WAL97}. Applications of these
``hyperpolarized" noble
gases in  medical imaging \cite{ALB94} and fundamental physics\cite{CHU94}
 are of
considerable current interest.  Nevertheless, the basic physical
limits to the efficiency of the spin-exchange process are still not
fully understood. Under typical operating conditions of
Rb-$^3$He spin exchange polarizers (saturated Rb vapor at
200$^\circ$C  in 8 amagat of $^3$He) about 45\% of the
Rb spin depolarization is due to Rb-Rb collisions \cite{BAR98}.
Here we report conclusive experimental evidence that much of
the spin relaxation in dense alkali-metal
vapors comes from the spin-axis interaction in triplet dimers.  Our measured spin-axis coupling strengths also
impact the field of ultracold collisions, where the spin-axis interaction causes trap-loss and affects the widths
and positions of Feshbach resonances\cite{Leo,Leo2}.

In 1960, Anderson {\it et al.} \cite{AND59} showed that
the dominant effect of binary collisions between
alkali-metal atoms is exchange of the electron spins.
%This
%process conserves the total spin angular momentum of the
%colliding atoms so the spins relax to a spin-temperature
%equilibrium  of spin temperature $\beta^{-1}$, where the
%relative population $\rho_{Fm}$ of atoms in
%a Zeeman sublevel $|Fm\rangle$ with azimuthal quantum
%number $m$ is given by the Boltzmann factor $e^{-\beta m}$.
%Spin-exchange between alkali-metal atoms is usually the
%fastest spin relaxation mechanism involved in spin-exchange
%optical pumping.
In 1974 Gupta {\it et al.} \cite{GUP74}
showed that the spin angular momentum
of the alkali-metal atoms was freely exchanged with the nuclear
spin of the singlet dimers, for example, Rb$_2$ and Cs$_2$
(where nuclear quadrupole interactions cause some spin depolarization).
%Even though the dimer density is only $10^{-3}$ that of the monomers,
%the dimers cause significant spin relaxation for Rb$_2$ because of the
%nuclear electric quadrupole couplings. The predicted quadrupolar
%relaxation from singlet dimers is much too small to account for the
%observed relaxation in dense Rb vapors in He or N$_2$ at pressures
%above one atmosphere. For $^{133}$Cs, with its unusually small
%quadrupole moment, the predicted relaxation is negligible compared to
%observed spin relaxation at any He or N$_2$ pressure. The singlet dimers
%are less important for Cs because of the unusually small quadrupole
%moment of
%$^{133}$Cs.
In 1980, Bhaskar {\it et al.} \cite{BHA80} showed that  Cs-Cs
interactions destroy spin at about 1\% of of the spin-exchange
rate.  The rate coefficient
for Cs-induced relaxation showed little dependence
on buffer gas pressure or species. Relaxation
mechanisms in  singlet or triplet  dimers
were expected to depend strongly on pressure. Consequently,
Bhaskar {\it et al.} \cite{BHA80} proposed that the relaxation was due
to binary collisions between alkali-metal atoms. They
pointed out that although there should be a spin-rotation
interaction,
$V_{\rm sr}=\gamma {\bf N}\cdot{\bf S}$ in the triplet dimers --
similar to that postulated
by Bernheim \cite{BER62}
% as the cause of spin-relaxation due to
%collisions between alkali-metal atoms and
%diamagnetic buffer gas atoms like He and N$_2$
-- the
coupling coefficient $\gamma$ was probably
too small to account for relaxation in binary
collisions.
%Their estimates were
%based on the experimental studies by Bouchiat \cite{BOU72a} and her
%collaborators of the magnitude of the spin-rotation coupling
%coefficient $\gamma$ for Rb paired with Ar, Kr and Xe in van der
%Waals molecules.
Instead, they suggested that the source of the relaxation
was a spin-axis interaction of the classic form \cite{HEB36}
\begin{equation}
V_{\rm sa} = {2\lambda\over 3 }{\bf S} \cdot (3\mbox{\boldmath{$\zeta$}}
\mbox{\boldmath{$\zeta$}}-
{\bf 1}) \cdot {\bf S}.
\end{equation} %  \eqno (1) $$
Here ${\bf S}={\bf S}_1+{\bf S}_2$ is the total electron spin of
the colliding pair, and \mbox{\boldmath{$\zeta$}} is a unit vector
along the direction
of the internuclear axis.  The  spin-axis coupling coefficient $\lambda=\lambda(R)$ arises from both the 
interaction energy of the electrons' magnetic dipole moments and the spin-orbit interaction in second
order\cite{Mies}.
 
%Little was
%known about the spin-axis coupling coefficient $\lambda=\lambda(R)$ and
%its dependence on the internuclear separation $R$, so it
%seemed possible that it could  be large enough (a few cm$^{-1}$) to
%account for the observed relaxation resulted from binary
%collisions of alkali-metal atoms in the triplet electronic
%spin state.

%Measurements of spin relaxation in dense Rb and K vapors by
%Knize \cite{KNI89} in 1989 showed that, as in the case of Cs, there
%was a contribution to the
%relaxation rate proportional to the density of the alkali-metal atoms,
%and almost independent of the pressure of the He or N$_2$ buffer gas.

Because Knize's \cite{KNI89}
measurements also showed little dependence of the Rb-induced
relaxation on He or N$_2$ pressure, it came as a surprise when in 1998,
Kadlecek,
Anderson and Walker \cite{KAD98} discovered that the  relaxation could be
substantially suppressed by
externally applied magnetic fields of a few thousand Gauss.
If the interaction were due to pairs of alkali-metal atoms
interacting on the triplet potential curve,  this field dependence sets a
lower limit on the correlation time of 40
ps,  the inverse of the
electron Larmor frequency at the decoupling field and a time much
longer than the duration of a binary collision.
Similar magnetic decoupling curves for K and
Cs vapors were soon observed in our laboratories at  Wisconsin and Princeton.

Although we still do not understand the
relaxation at He or N$_2$ pressures of an atmosphere
or more, recent low-pressure experiments leave no doubt
that an important part of the spin relaxation in dense
alkali-metal vapors comes from the spin-axis interaction (1),
acting in triplet dimer molecules -- even though the triplet
dimer density is no more than $10^{-6}$ that of the monomers.
The key experimental observation is the existence of resonances
in the magnetic decoupling curves, resonances which are
predicted from the spin-axis interaction (1), and which cannot
be produced by the spin-rotation interaction or by spin-1/2 species such as
trimer molecules.

A representative arrangement for measuring relaxation
transients in Rb is shown at the top of Fig.~1.
The alkali vapor and N$_2$ buffer gas  were contained in 1 inch spherical
glass bulbs that were heated in an %aluminum
oven placed between the pole faces of a 10 kG electromagnet.
We  measured the
number density of Rb atoms using
Faraday rotation \cite{WU86} of a tunable diode laser \cite{MAC92}.

\begin{figure}[htb]\vspace*{4.0 in}\includegraphics{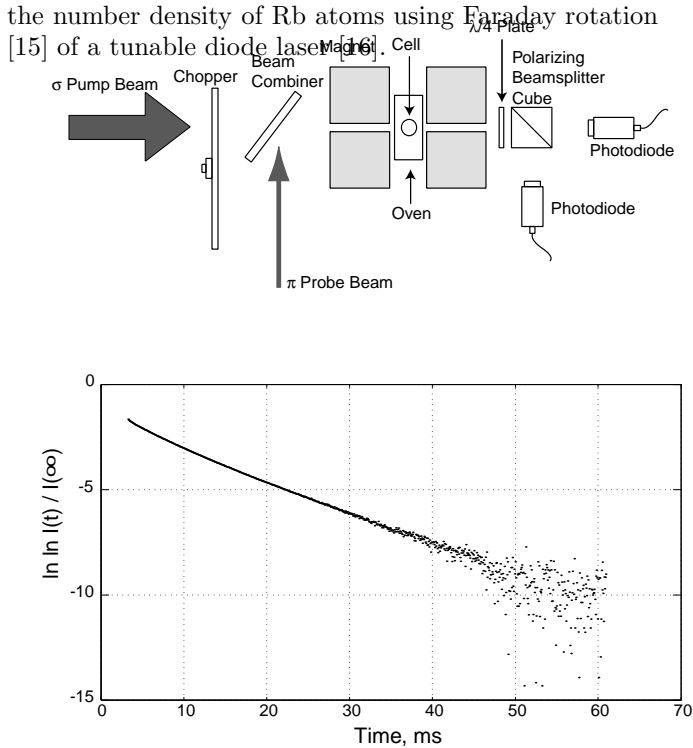}
\caption{
 {\it Top}. Experimental apparatus for measuring the
spin relaxtion of Rb vapor. {\it Bottom}.
Representative transient decay curve obtained with the apparatus;
at late-times the transient decay is characterized by a single exponential
time constant.}
\end{figure}

An intense pulse of 7947 \AA\
circularly polarized pump laser radiation built up the
spin polarization for a short (10 ms) interval. The pump
laser was then blocked by the chopper wheel, and photodiodes monitored the
transmission of a weak linearly polarized probe laser through the cell. The
probe
beam was resolved into its two circular components, giving positive and
negative helicity signals $V_{\pm}(t)$ for sampling times $t$.
The signal ratio $I(t)=V_+(t)/V_-(t)$
greatly suppressed  probe-laser noise.
The difference in the attenuation coefficients
of the two circular components of the probe laser is proportional
to the spin polarization $P$ (for $P \ll 1$) of the vapor, so $\ln I(t)=
a+bP(t)$, where $a=\ln I(\infty)$ and $b$
are constants. Fig. 1 shows a typical decay transient $\ln\ln
I(t)/I(\infty)$. For early times, the decay is not
exponential since the faster eigenmodes \cite{HAP72} of spin relaxation are
still contributing to the polarization $P$. However,
at late times there is a time interval, several $e$-foldings in length,
over which $\ln\ln I(t)/I(\infty)$ is nearly a straight line with
a slope $-\gamma$. Thus, when the  polarization has decayed
for a sufficiently long time, it is characterized by a single,
exponential {\it late-time decay rate} $\gamma$.

%The decay rates $\gamma$ were measured as a function of applied
%longitudinal magnetic field $B$. The magnetic field was measured
%with a Hall effect probe, which was independently calibrated with
%proton magnetic resonance.
%We studied the spin relaxation rate as
%a function of magnetic field for different alkali number densities,
%and for different third body densities.

The late-time decay rates of polarized Cs vapor in N$_2$  gas were
measured as described previously \cite{KAD98}, using Faraday rotation of
linearly polarized light both to
measure the Cs density and to monitor the spin-polarization.

Representative measurements of $\gamma$ in $^{87}$Rb are shown in
Fig. 2(a), where one can recognize 3 well-resolved resonances on
the magnetic decoupling curve at magnetic fields of approximately
900, 1500 and 2100 Gauss. Similar representative data for
$^{133}$Cs are shown in Fig. 3, where the theoretical curve has
7 poorly resolved resonances, which nevertheless provide a good
fit to the experimental data.

\begin{figure}[tbh]\vspace*{4.4 in}\includegraphics{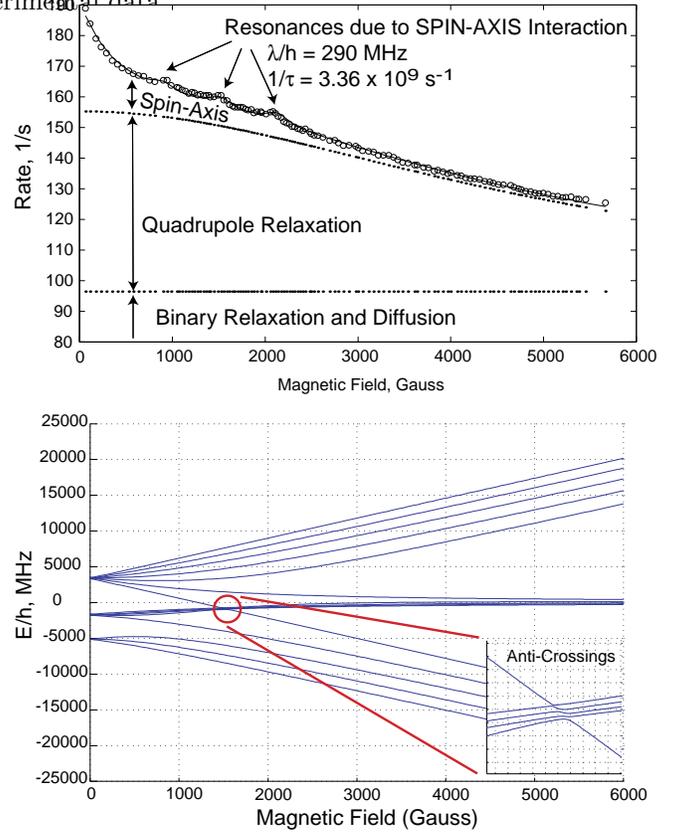}
\caption{ {\it Top}. Late-time decay
rate of $^{87}$Rb versus magnetic
field $B$. Resonant enhancements of the decay rate are observed
at the fields $B_1=915$ G, $B_2=1526$ G and $B_3=2136$ G, in agreement with
the predicted values
$B_I = A(2I+1)/4 g_S \mu_B$. The cell contained 0.061 amagat of
N$_2$ gas. %The monomer number density [Rb], as determined
%by Faraday rotation, is indicated on the Figure.
The cell temperature
was 220 C. The solid line is a theoretical fit assuming
a field-independent contribution to $\gamma$ of
$97$ sec$^{-1}$ from binary collisions and
diffusion to the cell walls, a field-dependent
contribution from quadrupole interactions in singlet dimers,
calculated as outlined in Ref.~\protect\cite{GUP74}, and a contribution
from a spin-axis
interaction in triplet dimers, calculated according
to Eq. (\protect\ref{kapeq}).
 {\it Bottom}. Energy levels for a representative spin space
with $I=2$. The level anticrossings are responsible for
the resonances in the magnetic decoupling curves.}\end{figure}

Also shown in Fig. 2(b)  are representative
energies of the spin sublevels of the triplet dimer molecules
of $^{87}$Rb
%and $^{133}$Cs
for a value of the total nuclear
spin quantum number $I=2$.
% and $I=7$ respectively.
The resonances in the relaxation rate occur at the
same magnetic fields as anticrossings in the energy level diagrams. The
anticrossings are the result of the spin-axis interaction (1),
acting in a total spin Hamiltonian for  homonuclear triplet dimers of
the form
\begin{equation}
  H ={A\over 2}{\bf S}\cdot{\bf I }+g_s\mu_B B S_z
  - {\lambda\over 3 }{\bf S} \cdot
  (3{\bf nn}-
  {\bf 1}) \cdot {\bf S}. %\eqno (2)$$
  \label{saham}
\end{equation}
Here $A$ is the magnetic-dipole hyperfine coupling coefficient for
the free alkali atoms ($A(^{87}\Rb)/h =3,417.4 $ MHz and
$A(^{133}\Cs)/h =  2,298.2 $ MHz).
%By using the free-atom
%values for $A$, we are assuming that the triplet dimers are
%so loosely bound that the magnetic fields produced by the two,
%parallel electrons at the two nuclei are nearly the same at the
%fields the electrons would produce if the atoms were not bound.
The operator for the total nuclear spin, ${\bf I} = {\bf I}_1+
{\bf I}_2$ is the sum of the nuclear spin operators ${\bf I}_1$
and ${\bf I}_2$ for the two atoms.
The operator ${\bf I}\cdot {\bf I}$ commutes with (2), and for homonuclear dimers its
eigenvalues $I(I+1)$ can have $I=0,1,2,\ldots,2I_m$, where $I_m=I_1=I_2$
is the nuclear spin quantum number of a monomer.
The Hamiltonian (2) can therefore be diagonalized in a spin subspace
$I$ of dimension $3(2I+1)$ to find eigenfunctions $|i\rangle=|Ii\rangle$
and
eigenvalues $E_{i}=E_{Ii}$.
For $\lambda=0$  there is a curious degeneracy in the spin subspace $I\ne0$,
with $2I+1$ sublevels crossing at the magnetic field
$B_I = A(2I+1)/4 g_S \mu_B$.
%The second term in (2) is the Zeeman interaction of the dimer
%electrons with the externally applied magnetic field $B$ which
%defines the $z$ axis of the coordinate system.
The last term in
(\ref{saham}) is the spin-axis interaction (1), where we assume that the
rotational angular momentum ${\bf N}$ of the dimers is large enough
that we may treat it as a classical vector, pointing along
the unit vector ${\bf n}={\bf N}/N$, and make the
replacement $(3\mbox{\boldmath{$\zeta$}}
\mbox{\boldmath{$\zeta$}}- {\bf 1}) \to -(3{\bf n}{\bf n}-{\bf
1})/2$.

We estimate the relaxation caused by the interaction (2)
with the following simple model.
Let the total longitudinal spin operator of the
triplet dimer be $G_z=S_z+I_z$. When triplet dimers are formed
in dense alkali-metal vapors with low spin polarization, the spin
density matrix of the dimers will be very nearly $\rho(0) =
e^{\beta G_z}/Z$, where  $\beta\approx 4P\ll 1$ is the spin-temperature
parameter \cite{AND59}. The partition function is very nearly
$Z=3(2I_m+1)^2$. The spin of the dimer at the instant of
formation is therefore $\langle G_{z0}\rangle = 2(I_m^2+I_m+1)\beta/3$,
where we have neglected non-linear terms in the small parameter $\beta$.
The mean change in the density matrix by the time it breaks up again is
$\Delta \rho =\langle U\rho(0)U^{-1}\rangle -\rho(0)$, where the time
evolution operator is $U=e^{-iHt/\hbar}$, and  the angle brackets
$\langle
\cdots \rangle$ denote an average over all directions
of $\bf n$, and also an average over the probability
$e^{-t/\tau}dt/\tau$ that the dimer will break up in the time
interval between $t$ and $t+dt$. The collisional breakup rate of the
dimers is $1/\tau$.  The mean change in the dimer spin is therefore
$\langle \Delta G_z\rangle = \Tr \{ G_z \Delta \rho \}=-W\langle
G_{z0}\rangle$, where the spin-destruction probability is
\begin{equation}
W = 2\sum_{ij}{|\langle i|G_z|j\rangle|^2(\omega_{ij}\tau)^2
\over (2I_m+1)^2[(2I_m+1)^2+3][1+(\omega_{ij}\tau)^2]}, \label{weq}
%\eqno (3)$$
\end{equation}
and the Bohr frequencies are $\omega_{ij} = (E_i-E_j)/\hbar$.
An average over the directions of ${\bf n}$
is to be understood in (\ref{weq}).
Almost all of the spin is carried by monomers of
number density [Ak] ({\it e.g.}, [Ak]=[Rb] or [Ak]=[Cs]),
which have a spin density per unit
volume $[\Ak]\beta (4I_m^2+4I_m+3)/12$. The spin loss rate per
unit volume from triplet dimers of number density $[\Ak_2]$
is $[\Ak_2]\langle \Delta G_z\rangle/\tau$, from which one
obtains the relaxation equation
%
%\begin{equation}
$-(d/ dt)\ln{\beta} = \gamma = \kappa[\Ak]$, %  \eqno (4)$$
%\end{equation}
%
where the rate coefficient is
\begin{equation}
\kappa =2\left[{(2I_m+1)^2+3\over (2I_m+1)^2+2}\right] {W{\cal{K}}
\over \tau}. \label{kapeq}$$
\end{equation}
The chemical equilibrium coefficient of the triplet dimers is
%
%\begin{equation}
${\cal K}={[\Ak_2]/ [\Ak]^2}$. %\eqno (6)$$
%\end{equation}
%
We have used theoretical triplet dimer potential curves of
Krauss and Stevens \cite{KRA90} to estimate ${\cal K}({\rm Rb}) =330 $
\AA$^3$  and ${\cal K}({\rm Cs}) = 650$
\AA$^3$ at representative temperatures of 220$^\circ$C and 180$^\circ$C
respectively.

\begin{figure}[tbh]\vspace*{3.2 in}\includegraphics{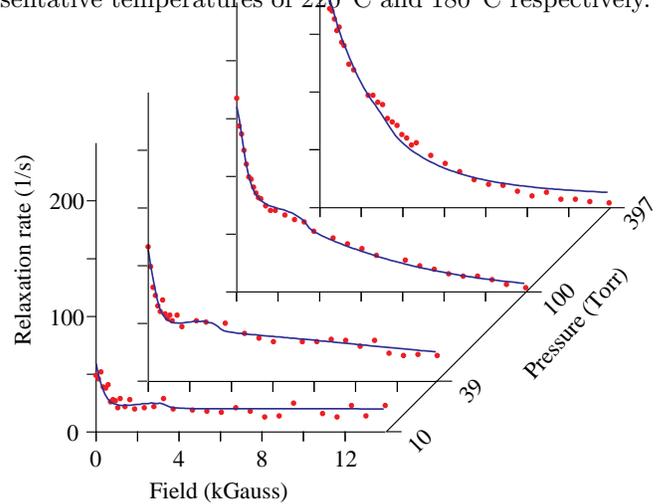}
%\special{psfile=thirdtry.epsf hscale=40 vscale=40}
\caption{ Late-time field-dependent decay rate of
$^{133}$Cs versus magnetic field and pressure at 180$^\circ$C,
[Cs]$=6.6\times 10^{14}$ cm$^{-3}$. The cell
temperature was 177 C. The fits to the data use the reorientation model
with negative $\lambda$.}
\end{figure}

%%%%%%
%If the spin-axis coupling coefficient $\lambda$ vanishes in (2),
%one can verify that $W=0$ in (3), since the energy eigenstates
%$|i\rangle$ are simultaneously eigenstates of $G_z$ with azimuthal
%quantum number $m$. Furthermore, there is a curious degeneracy
%in the spin subspace $K\ne 0$, and one finds that $2K+1$ sublevels
%cross at the magnetic field
%
%\begin{equation}
%B_K={ A(2K+1)\over 4g_S\mu_B }. %\eqno (7)$$
%\end{equation}
%
%Although one might expect a resonance at the crossing fields
%(7), since the Bohr frequencies of crossing levels vanish
%and the factor $1+\omega_{ij}\tau$ attains its minimum
%value of $1$ in the denominator of (3),
%$W$ remains zero near the level crossing fields (7) because
%the corresponding matrix elements $\langle i|G_z|j\rangle$ in the
%numerator of (3) vanish, a result of $m$ being a good quantum
%number.
%%%%

Non-vanishing values of $\lambda$ lift the degeneracy at the field
$B_I$ and convert the level crossings into anticrossings, while making
the corresponding matrix element $\langle i|G_z|j\rangle$
non-zero, as is illustrated in Fig. 2. It is these anticrossings of the
energy levels that
are responsible for the resonances, most clearly visible in Fig. 2.
Remarkably, the spin-rotation interaction does not break the degeneracy so
the spin-axis interaction must
be responsible for the relaxation resonances.
Since there are $2I_m$ nonvanishing values of $I$ for atoms with the
nuclear spin quantum number $I_m$, we expect there to be $3$
resonaces for $^{87}$Rb with $I_m=3/2$, as observed, and $7$
resonances, at fields between $600$ G and $3100$ G for $^{133}$Cs
with $I_m=7/2$, which fits well the observed field dependence shown in Fig~3.

We  fit the experimental data to
theoretical curves that include a field-independent part due to
diffusion to the cells walls and binary collisions, a field-dependent
nuclear quadrupole relaxation in singlet
dimers (negligible for Cs), and relaxation due to the spin-axis interaction
in triplet
dimers. We expect a small portion of the relaxation in triplet dimers
to arise from  the spin-rotation interaction, and we may estimate the
magnitude of this  relaxation mechanism by assuming that the
spin-rotation
interaction of Rb-Rb triplet dimers is the same as that that of Rb-Kr,
and that the spin-rotation interaction for Cs-Cs is the same as
that for Rb-Xe, both of which are experimentally known from the
work of  Bouchiat {\it et al.} \cite{BOU72b}. These estimates should be fairly reliable since the
 spin-orbit interaction that produces the spin-rotation coupling \cite{WAL97} varies only slightly from Rb to Kr
or Cs to Xe. Using these estimates for
$\gamma {\bf N}\cdot {\bf S}$, we find that
the addition of the spin-rotation interaction causes a
few percent change in the predicted relaxation rates.

In the model described above, the relaxation due to triplet dimers
dependends on two parameters: $\lambda$ and the
molecular lifetime $\tau$, assuming the theoretical values for $\cal K$ are
accurate. The molecular lifetime should
be inversely proportional to pressure, making it natural to introduce a
cross-section
$\sigma=(\tau[\rm{N}_2]\overline{v})^{-1}$. Our Rb data is well-described
over the limited pressure range from 50 Torr
to 300 Torr by the values $|\lambda|=290$ MHz, and $\sigma=290$ \AA$^2$.
The latter is somewhat larger than the
typical 150 \AA$^2$ breakup cross sections deduced from magnetic decoupling
studies of RbKr\cite{BOU72b}. The sign of
$\lambda$ cannot be determined from the data.

Since there is little contribution to Cs relaxation from singlet molecules,
relaxation from Cs-Cs triplet molecules
can be observed over a wider range of pressures. Fig.~3 shows a sampling of
our spin-relaxation data for
 Cs  as a function of magnetic field at a variety of N$_2$ pressures.   In this case the data are
well-described by $\lambda=-2.9$ GHz, $\sigma=98$ \AA$^2$, $
{\cal K}=350$ \AA$^3$, or, assuming the
opposite sign of $\lambda$, $\lambda=3.4 $ GHz, $\sigma=82$ \AA$^2$, ${\cal
K}=307$ \AA$^3$. The deduced
values of $\cal K$ are about a factor of 2 smaller than predicted from the
potential curves (which are believed to be
fairly reliable).  We have also fit the data with a model that allows for
collisional reorientation of the triplet
molecules without breakup, making the assumption that a spin-temperature
distribution describes the dimer
density matrix after reorientation. In this case
$\tau$ in Eq.~\ref{weq} is replaced by the coherence time
$\tau_c$, and
$\overline{W}=W \tau/(\tau_c+(\tau-\tau_c) W)$ replaces $W$ in
Eq.~\ref{kapeq}.  In this case we obtain
$\lambda=-2.13 $ GHz, $\sigma_c=81$ \AA$^2$, $\sigma=42$
\AA$^2$, ${\cal
K}=584$ \AA$^3$, or
$\lambda=2.71$ GHz, $\sigma_c=93$ \AA$^2$,  $\sigma=55$ \AA$^2$, ${\cal
K}=431$ \AA$^3$.
 We emphasize that single values of $\lambda$ and $\sigma_c$ accurately
represent the magnetic field dependences at 9
pressures (not all shown in Fig~3), covering nearly two decades in pressure.

Spin-axis coupling in Cs triplet dimers is also of current interest for
studies of ultracold collisions\cite{Leo}
since it produces loss of spin-polarized atoms in magnetic traps, and
affects the widths and positions of Feshbach
resonances in Cs collisions\cite{Leo2}. The magnitude and sign of
$\lambda$, as well as its dependence on interatomic
separation $R$ were predicted by Mies {\it et al.}\cite{Mies}. Previous
analysis\cite{Leo} led to the conclusion that
in order to agree with experiment the second-order spin-orbit calculated by
Mies {\it et al.} needed to be
scaled by a factor of
$S_C=4.0\pm 0.5$, recently updated to $3.2\pm 0.5$\cite{Leo2}.  We find, by
averaging $\lambda(R)$ over the
ro-vibrational levels of the triplet state using the Krauss and Stevens
potentials\cite{KRA90}, scaling factors
of $S_C=12$ for Rb and
$S_C=3.6$ for Cs are needed to reproduce our data.

In conclusion, the resonances observed in the
magnetic decoupling curves of Rb, coupled with the observed pressure and
field dependences of Cs, prove that
a major contribution to the  spin relaxation
of hot, dense alkali-metal vapors is caused by the
spin-axis interaction in triplet dimers. Knowing
this key piece of the physics should
help to unravel the still-puzzling independence of
the relaxation rates on He and N$_2$ pressures
of an atmosphere or more \cite{KAD98}.

Support for this research came from  NSF (Wisconsin \& Princeton), and  NIH
(Princeton).


\begin{references}

\bibitem{WAL97} T. G. Walker and W. Happer, Rev. Mod. Phys. {\bf 69}, 629
(1997).

\bibitem{ALB94} %1.
M. S. Albert, G. D. Cates, B. Driehuys, W. Happer, B.
Saam, C. S. Springer and A. Wishnia, Nature, {\bf 370}, 199
(1994).



\bibitem{CHU94} %2.
T. E. Chupp, R. J. Hoare, R. L. Walsworth, and Bo Wu,
Phys. Rev. Lett., {\bf 72},  2363 (1994).


\bibitem{BAR98} A. Ben-Amar Baranga, S. Appelt, M. V. Romalis, C. J.
Erickson, A. R. Young, G. D. Cates, and W.
Happer, Phys. Rev. Lett.
{\bf 80}, 2801 (1998).

\bibitem{Leo} P. Leo, E. Tiesinga, P. Julienne, D. Walter, S. Kadlecek, and
T. Walker, Phys. Rev. Lett. {\bf 81}, 1389 (1998)

\bibitem{Leo2} P. Leo, C. Williams, and P. Julienne, preprint.


\bibitem{AND59} %4.
L. W. Anderson, F. M. Pipkin and J. C. Baird, Phys. Rev.
{\bf 116}, 87 (1959).


\bibitem{GUP74} %5.
R. Gupta, W. Happer, G. Moe and W. Park, Phys. Rev. Lett.
{\bf
32} 574 (1974); Note that in Eq. 7 of this paper, the numerical
coefficient $3/40$ should have been $3/160$.


\bibitem{BHA80} %6.
N. D. Bhaskar, J. Pietras, J. Camparo, and W. Happer, Phys. Rev. Lett.,
{\bf 44}, p. 930
(1980).


\bibitem{BER62} %7.
R. A. Bernheim, J. Chem. Phys. {\bf 36}, 135 (1962).


%\bibitem{BOU72a} %8.
%M. A. Bouchiat, J. Brossel and J. Pottier, J. Chem.
%Phys. {\bf 56}, 3703 (1972).
%\smallskip


\bibitem{HEB36} %9.
M. H. Hebb, Phys. Rev. {\bf 49}, p. 610 (1936); C. H.
Townes and
A. L. Schawlow, {\it Microwave Spectroscopy,}\ p. 182, Dover
Publications,
New York (1975).

\bibitem{Mies}  F. H. Mies, C. Williams, P. Julienne, and M. Krauss, J.
Res. Natl. Inst. Stand. Technol. {\bf 101},
521 (1996).


\bibitem{KNI89} %10.
R. Knize, Phys. Rev. A {\bf 40}, 6219 (1989).


\bibitem{KAD98} %11.
S. Kadlecek, L. W. Anderson and T. G. Walker, Phys. Rev.
Lett. {\bf 80}, 5512 (1998).


\bibitem{WU86} %12.
Z. Wu, M. Kitano, W. Happer, M. Hou, and J. Daniels,
Applied Optics {\bf 25}, 4483  (1986)


\bibitem{MAC92} %13.
K. B. MacAdam, A. Steinbach, and C. Wieman,
Am. J. Phys. {\bf 60}, 1098  (1992).

\bibitem{HAP72} W. Happer, Rev. Mod. Phys. {\bf 44}, 169 (1972).

%\bibitem{NAG78}
%W. Nagourney, W. Happer, A. Lurio, Phys. Rev. A {\bf 17},
%1394 (1978).



\bibitem{KRA90} %15.
M. Krauss and W. J. Stevens, J. Chem. Phys. {\bf
93}, 4236 (1990).


\bibitem{BOU72b} %16.
M. A. Bouchiat, J. Brossel, and L. C. Pottier,
J. Chem. Phys. {\bf 56}, 3703 (1972).


\end{references}
\end{document}